\documentclass[9pt]{article}

\usepackage[utf8]{inputenc}
\usepackage[T1]{fontenc}
\usepackage{amsmath, amssymb, amsthm}
\usepackage{graphicx}
\usepackage{booktabs}
\usepackage{float}
\usepackage{hyperref}
\usepackage{microtype}
\usepackage{xcolor}
\usepackage{multicol}
\usepackage{lipsum} 
\usepackage{cite}

\usepackage[top=1.8cm, bottom=2.1cm, left=1.75cm, right=1.75cm]{geometry}

\usepackage{authblk}

\renewenvironment{abstract}
{\centerline{\bfseries Abstract}\par\noindent\ignorespaces}
{\par}

\begin{document}


{\fontfamily{ptm}\LARGE \bfseries \noindent
Bubble damping of non-stationary oscillatory flow stabilization in microfluidic systems}\\[0.5cm]

\vspace{-5mm}
{\noindent
Andreu Benavent-Claró\textsuperscript{1,*}\\[0.2cm]

\vspace{-4mm}
{\footnotesize \it \noindent
\textsuperscript{1}Condensed Matter Physics Department, Physics Faculty, University of Barcelona, Spain.\\
\textsuperscript{*}\texttt{abenavent@ub.edu}
}\\[0.8cm]
}

\vspace{-8mm}
\noindent
\begin{minipage}[t]{\textwidth}
\centering
\begin{minipage}[t]{0.8\textwidth} 
\begin{abstract}
The inherent instability of oscillatory flows presents a significant challenge in microfluidics, impairing performance in different applications from particle detachemnt to organs-on-a-chip. Trapped air inside a microfluidic system passively dampens these fluctuations because of the compressible nature of air. However, a foundational theoretical model that describes this effect has remained elusive. Here, a first-principles model that fully characterizes the effects of a trapped air volume in oscillatory microfluidic flow is derived. The model identifies a dimensionless product (\(\tau \omega\)) as the governing parameter, unifying the interplay between air compressibility and fluidic resistance. It precisely predicts the volume displacement dynamics of the liquid front, which compared with the original flow, it presents amplitude reduction, phase shift, and transient drift. The theoretical framework was validated with different experiments across a broad range of conditions. This work transforms trapped air from a source of unpredictability into a powerful, predictable element for tailoring oscillatory flow stability, providing a rigorous design tool for microfluidic systems.
\end{abstract}

\end{minipage}
\end{minipage}

\vspace{5mm}

\begin{multicols}{2}

\section{Introduction}

Microfluidic systems are foundational to advancements in diverse fields such as point-of-care diagnostics, surface cleaning, fluid transport and sophisticated chemical synthesis \cite{whitesides2006, sia2008, danku2022, browne2021}. A common requirement across these applications is the precise manipulation of fluids within confined geometries, where the stability and predictability of flow are paramount. Oscillatory flows, in particular, are employed for tasks including enhanced mixing, particle manipulation, and mimicking physiological conditions \cite{ottino1989, di2006, young2007, hur2011,simoes2005}.

The presence of a trapped air volume is a common and often problematic feature in such systems. It arises inadvertently in devices like syringes or can be deliberately incorporated, for example, in certain pressure sensors that use an air bubble as an interface to isolate the sensor from the liquid \cite{srivastava}. The inherent compliance of the air leads to complex, damped responses that are difficult to predict or control, posing a significant challenge for device reliability \cite{vobecka2023}. This has traditionally forced researchers and engineers to treat air as a nuisance, leading to designs that attempt to eliminate it rather than understand it.

When air bubbles are present on the microfluidic system, the motion does not propagate instantaneously through the air due to its compressibility \cite{lyubimova2023}. If the motion applied is non-stationary as it is an oscillatory flow, the compressibility of air and the fluidic resistance of the liquid makes that the motion propagation damped and phase-shifted.

Paradoxically, this same compressibility can be harnessed for beneficial damping \cite{zhang2022}. A trapped air volume acts as a fluidic capacitor, passively filtering out high-frequency fluctuations and stabilizing the liquid flow \cite{kang}. Although it has not been extensively researched as a primary phenomenon, this damping effect has been observed incidentally and used empirically in specific contexts \cite{lee,Hess}. Studies have noted that the motion transmitted to a liquid front is smoother than the plunger's own motion due to the compressible effect of the air compartment present between the plunger and the liquid\cite{lee}. Others have highlighted the general benefits of air-based damping, identifying key empirical properties like amplitude reduction \cite{srivastava, Hess}, and some have even characterized an empirical time constant governing these dynamics \cite{Araci}. The deliberate use of compliance has therefore been proposed as a method to dampen hydrodynamic fluctuations \cite{iyer2015experimental, doh2009passive}, with recent work exploring the nonlinear responses produced by pneumatic capacitance in microfluidic oscillators.

Despite these valuable observations, a comprehensive theoretical framework has remained absent. This phenomenon is often treated as a secondary issue in broader studies, resulting in predominantly empirical models. These approaches provide isolated, fitted relationships for specific phenomena such as amplitude reduction or phase shift but lack a first-principles foundation that provides a mathematical description for the liquid motion. Crucially, it cannot predict the full spatiotemporal response of the fluid front, leaving key dynamics like the transient drift of the mean position unexplained. The absence of a fundamental theory has forced designers to rely on ad-hoc solutions and trial-and-error, limiting the robust integration of compressible elements in microfluidic systems \cite{kalantarifard}.

In this work, this gap is addressed by deriving a unified mathematical model directly from fundamental physical principles. This approach, based on the physics of air compressibility and viscous flow resistance, yields a first-principles theoretical framework that moves beyond empirical correlation. This model provides a complete analytical description of the system dynamics, simultaneously predicting the amplitude reduction, phase shift, and the previously unexplained transient drift. It unifies these phenomena under a single dimensionless parameter, $\tau\omega$, whose constituents are derived from the system's fundamental geometry and material properties.

The applicability of this theory extends beyond the canonical example of a bubble in a syringe. It provides a general framework for understanding and designing microfluidic systems where a compressible volume as a trapped bubble, a flexible membrane, or a dedicated air chamber interacts with an oscillatory flow. By transforming air compliance from an unpredictable nuisance into a quantifiable design element, this work enables a new level of precision in controlling fluidic stability.

This model was validated with different experiments using a custom oscillatory syringe-pump system, demonstrating exceptional agreement between theory and data across a wide range of operating conditions. By providing a rigorous foundation that explains previously disparate empirical findings, this work establishes a predictive and general framework for harnessing air damping in oscillatory microfluidic systems.

\section{Mathematical Model}
\label{sec:model}

\begin{figure}[H]
    \centering
    \includegraphics[width=0.45\textwidth]{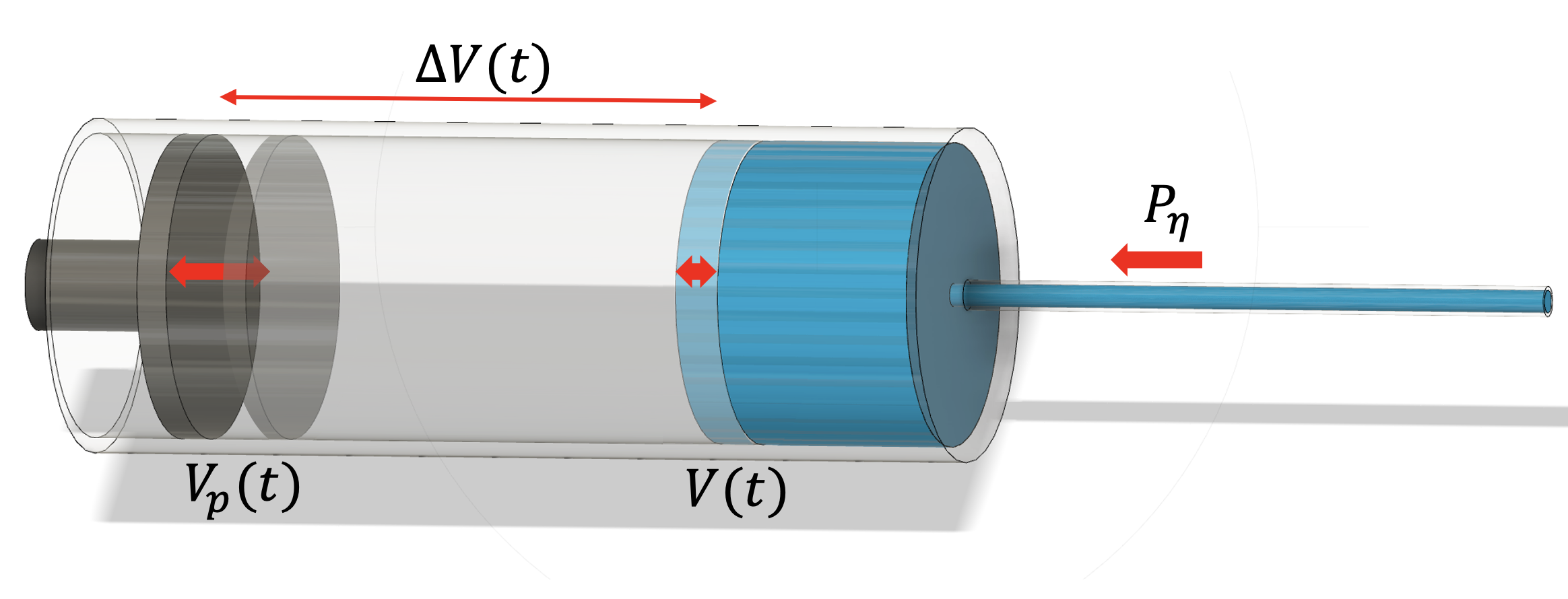}
    \caption{Schematic of the oscillatory flow system with a trapped air volume. The piston undergoes sinusoidal motion, compressing and expanding the air volume, which in turn drives the liquid through a narrow tube with significant fluidic resistance.}
    \label{Schematic}
\end{figure}

A mathematical model is developed to describe the dynamics of an oscillatory flow system containing a trapped air volume. This model aims to provide a comprehensive theoretical framework that captures the essential physics of how the compressibility of air affects the transmission of oscillatory motion from a piston to a liquid front. The system consists of a piston (plunger) moving within a syringe, which contains air acting as a compressible (elastic) fluid, and a liquid sample that behaves as a viscous fluid. The liquid exits the syringe through a narrow tube that provides significant fluidic resistance. The experimental setup is configured such that the liquid-air interface remains within the syringe at all times, ensuring that the air volume remains trapped and can act as a compliant element in the system. A schematic of the system is shown in Fig.~\ref{Schematic}.

The foundation of this model begins with the kinematics of the piston motion. The piston undergoes oscillatory motion, displacing a time-varying volume of fluid given by:
\begin{equation}
    V_p(t) = V_0 \sin(\omega t + \xi),
    \label{eq:V_piston}
\end{equation}
where $ V_0 $ is the amplitude of the piston motion, $ \omega $ is the angular frequency, and $ \xi $ is the initial phase. So, the initial condition is defined such that at $ t = 0 $, the piston displacement is $V_0\sin{\xi}$, meaning that the oscillation does not necessary starts from the equilibrium position.

The piston motion is transmitted to the liquid via the compressible air layer. Due to the compressibility of air, the volume displaced at the liquid-air front, denoted $ V(t) $, differs from the piston-displaced volume $ V_p(t) $. This difference arises because the air can store energy through compression and release it through expansion, effectively acting as a mechanical spring. The volumetric strain of the air is defined as:
\begin{equation}
    \gamma(t) = \frac{\Delta V(t)}{V_A} = \frac{V_p(t) - V(t)}{V_A},
    \label{eq:strain}
\end{equation}
where $ \Delta V(t) $ is the change in air volume (positive during expansion, negative during compression) and $ V_A $ is the initial volume of air. This strain definition follows the conventional approach for compressible fluids under small deformations.

Assuming Hookean behavior for the compressible air, which is valid for small strains and relatively slow compression rates, the restorative pressure is given by:
\begin{equation}
    P_K(t) = -K \gamma(t) = -\frac{K}{V_A} \left(V_p(t) - V(t)\right),
    \label{eq:Pk}
\end{equation}
where $ K $ is the bulk modulus of air. This pressure represents the elastic restoring force that the compressed air exerts on the liquid front.

Simultaneously, the pressure drop due to viscous resistance in the outlet tube is described by the Poiseuille equation, which assumes laminar, fully developed flow:
\begin{equation}
    P_\eta(t) = \mathcal{R} Q(t) = \frac{8 \eta l_t}{\pi r^4} \dot{V}(t),
    \label{eq:Peta}
\end{equation}
where $ \mathcal{R} $ is the fluidic resistance, $ Q(t) = \dot{V}(t) $ is the volumetric flow rate at the liquid-air front, $ \eta $ is the dynamic viscosity of the liquid, and $ l_t $ and $ r $ are the length and radius of the tube, respectively. This resistance term captures the energy dissipation due to viscous effects as the liquid flows through the narrow tube.

Applying a force balance on the liquid-air interface and neglecting inertial effects (valid for low frequencies) where acceleration terms are negligible compared to pressure and viscous forces, I obtain:
\begin{equation}
    P_K(t) + P_\eta(t) = 0,
    \label{eq:force_balance}
\end{equation}

Substituting Eqs.~\eqref{eq:Pk} and \eqref{eq:Peta} into Eq.~\eqref{eq:force_balance} yields the governing differential equation for the system:
\begin{equation}
    \dot{V}(t) = \frac{K \pi r^4}{8 l_t  \eta  V_A} \left( V_p(t) - V(t) \right).
    \label{eq:diff_eq}
\end{equation}

This first-order linear ordinary differential equation can be solved analytically. Substituting the piston motion from Eq.~\eqref{eq:V_piston} and solving with the initial condition $V(0) = V_0 sin(\xi)$, I obtain the following solution:
\begin{equation}
\begin{split}
    V(t) = \frac{V_0} {\tau^2 \omega^2 + 1} \left(\sin(\omega t+\xi)-\tau\omega\cos(\omega t+\xi)\right.\\
    \left.+\tau\omega\left(\cos(\xi)+\tau\omega\sin(\xi)\right)e^{-\frac{t}{\tau}}\right)
\end{split}
    \label{eq:V_solution_raw}
\end{equation}
where the time constant $ \tau $ is defined as:
\begin{equation}
    \tau = \mathcal{R} \mathcal{C} = \frac{8 l_t \eta V_A}{K \pi r^4},
    \label{eq:tau}
\end{equation}
and $ \mathcal{C} = V_A / K $ is the compliance of the air. This time constant represents the characteristic response time of the system and plays a crucial role in determining the dynamics.

Equation~\eqref{eq:V_solution_raw} can be simplified using the trigonometric identity $\sin(x)+b\cos(x)=\sqrt{b^2+1}sin(x-\arctan(b))$ to express the solution in terms of a single phase-shifted sine function, which provides clearer physical interpretation:
\begin{equation}
    V(t) = \frac{V_0}{\sqrt{ \tau^2 \omega^2 + 1 }} \left( \sin(\omega t+\xi-\phi)+\tau\omega\cos(\xi-\phi)e^{-\frac{t}{\tau}} \right),
    \label{eq:V_solution_final}
\end{equation}
where the phase shift $ \phi $ is given by:
\begin{equation}
    \phi = \arctan(\tau \omega).
    \label{eq:phase_shift}
\end{equation}

The solution consists of two distinct components with clear physical interpretations: a periodic term describing the oscillatory motion of the liquid front, and a transient term describing the exponential approach to a steady-state oscillation center. The amplitude of the front oscillation is:
\begin{equation}
    V_{\text{front}} = \frac{V_0}{\sqrt{\tau^2 \omega^2 + 1}},
    \label{eq:amplitude}
\end{equation}
which is always less than the piston amplitude $ A_p = V_0 $. The ratio of amplitudes is:
\begin{equation}
    \frac{V_{\text{front}}}{V_0} = \frac{1}{\sqrt{\tau^2 \omega^2 + 1}}.
    \label{eq:amplitude_ratio}
\end{equation}
This relationship shows that the amplitude reduction depends on the product $ \tau\omega $, which represents the ratio of the oscillation period to the system's response time.

The mean position about which the front oscillates evolves exponentially in time due to the transient term in Eq.~\eqref{eq:V_solution_final}, where I can see that the oscillation center shifts by:
\[
\frac{V_0 \tau\omega\cos(\xi-\phi)e^{-\frac{t}{\tau}}}{\sqrt{\tau^2 \omega^2 + 1}},
\]
approaching zero at long times. See that even if the plunger is centered ($\xi=0$) at initial time, center of the front oscillation also gets shifted, with the form $\frac{V_0 \tau\omega e^{-\frac{t}{\tau}}}{\tau^2 \omega^2 + 1}$, since $\cos(-\phi)=\frac{1}{\sqrt{\tau^2\omega^2+1}}$.

The flow rate of the liquid is found by differentiating Eq.~\eqref{eq:V_solution_final}:
\begin{equation}
    Q(t) = \frac{Q_0}{\sqrt{\tau^2 \omega^2 + 1}} \left( \cos(\omega t+\xi-\phi)-\cos(\xi-\phi)e{-\frac{t}{\tau}} \right),
    \label{eq:flow_rate}
\end{equation}
where $ Q_0 = V_0 \omega $. This expression reveals how the flow rate depends on both the oscillatory driving and the system's transient response.

Finally, the pressure driving the liquid flow can be expressed as:
\begin{equation}
    P_\eta(t) = \frac{V_p(t) - V(t)}{\mathcal{C}}
    \label{eq:Pliquid}
\end{equation}
This completes the mathematical description of the system, providing a comprehensive framework for understanding how a trapped air volume modifies the transmission of oscillatory motion in microfluidic systems.

The key insight from this model is that the product $ \tau\omega $ serves as a dimensionless parameter that governs the system's behavior. When $ \tau\omega \ll 1 $ (low frequency or small time constant), the system responds quasi-statically with minimal phase shift and amplitude reduction. When $ \tau\omega \gg 1 $ (high frequency or large time constant), significant damping and phase lag occur, effectively filtering the oscillatory motion. This theoretical framework provides the foundation for designing microfluidic systems with tailored oscillatory response characteristics.

\section{Methodology}

\subsection{Power source and pump mechanism}

\begin{figure} [H]
    \centering
    \includegraphics[width=0.39\textwidth]{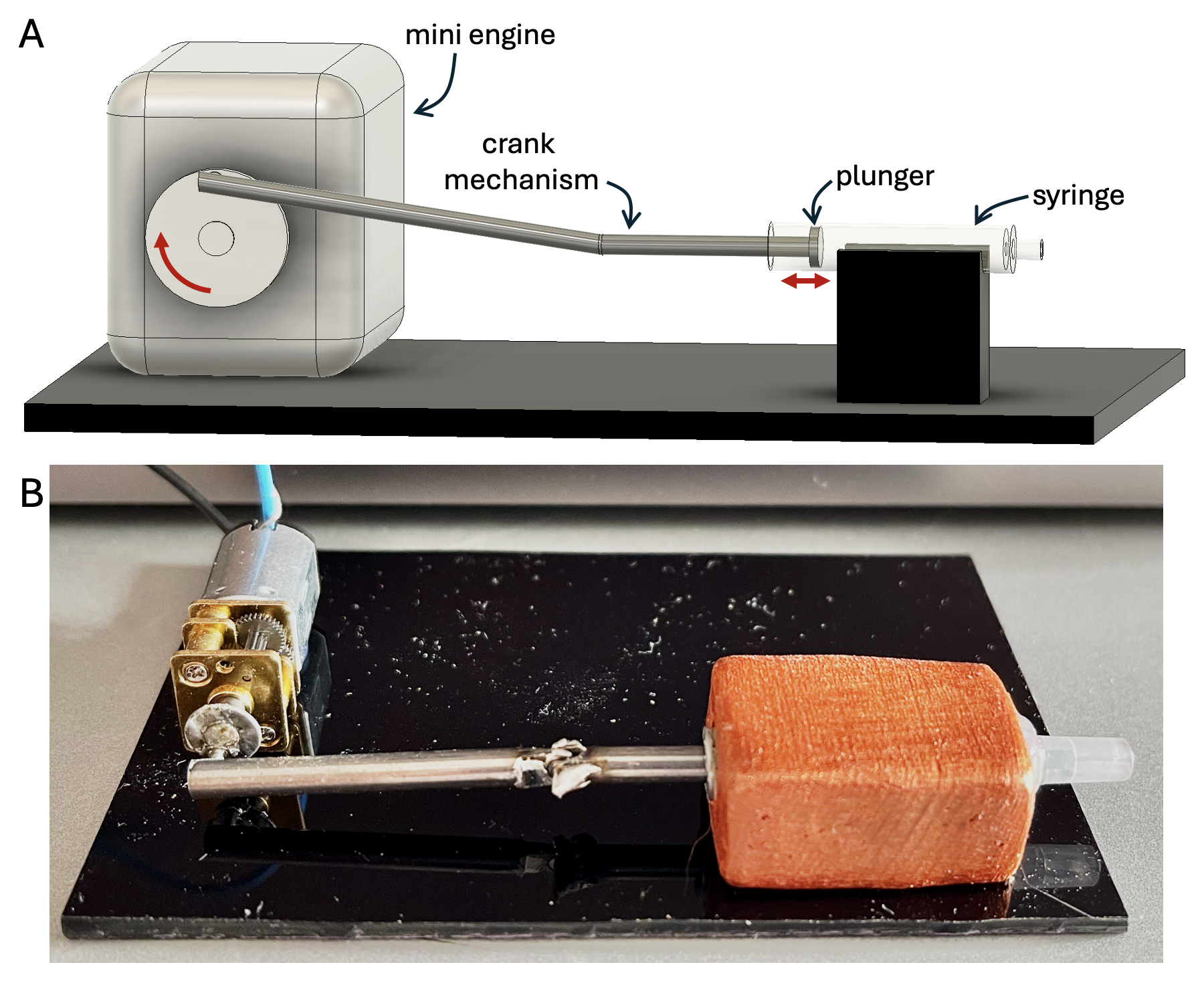}
    \caption{Custom oscillatory syringe pump. (A) Schematic diagram of the crank mechanism converting rotary motor motion to sinusoidal plunger displacement. (B) Photograph of the manually fabricated prototype showing the motor, reduction gear, connecting rods, and syringe assembly.}
    \label{fig:bomba}
\end{figure}

A custom oscillatory syringe pump was developed to generate sinusoidal motion of the plunger. The system employs a small electric motor equipped with a reduction gear to achieve angular velocities between 0.2 and 9 Hz. A variable resistor controls the motor voltage, enabling frequency adjustment across experiments. The rotary motion is converted to sinusoidal displacement via a crank mechanism consisting of two metal rods connected by bearings. One end connects to the motor while the other attaches to the plunger of a 1 mL syringe via thermofusion, with the syringe positioned to ensure alignment with the motor center. A washer welded to the piston and connected through a bearing minimizes friction while maintaining the required oscillation amplitude.

\begin{figure}
    \centering
    \includegraphics[width=0.62\textwidth]{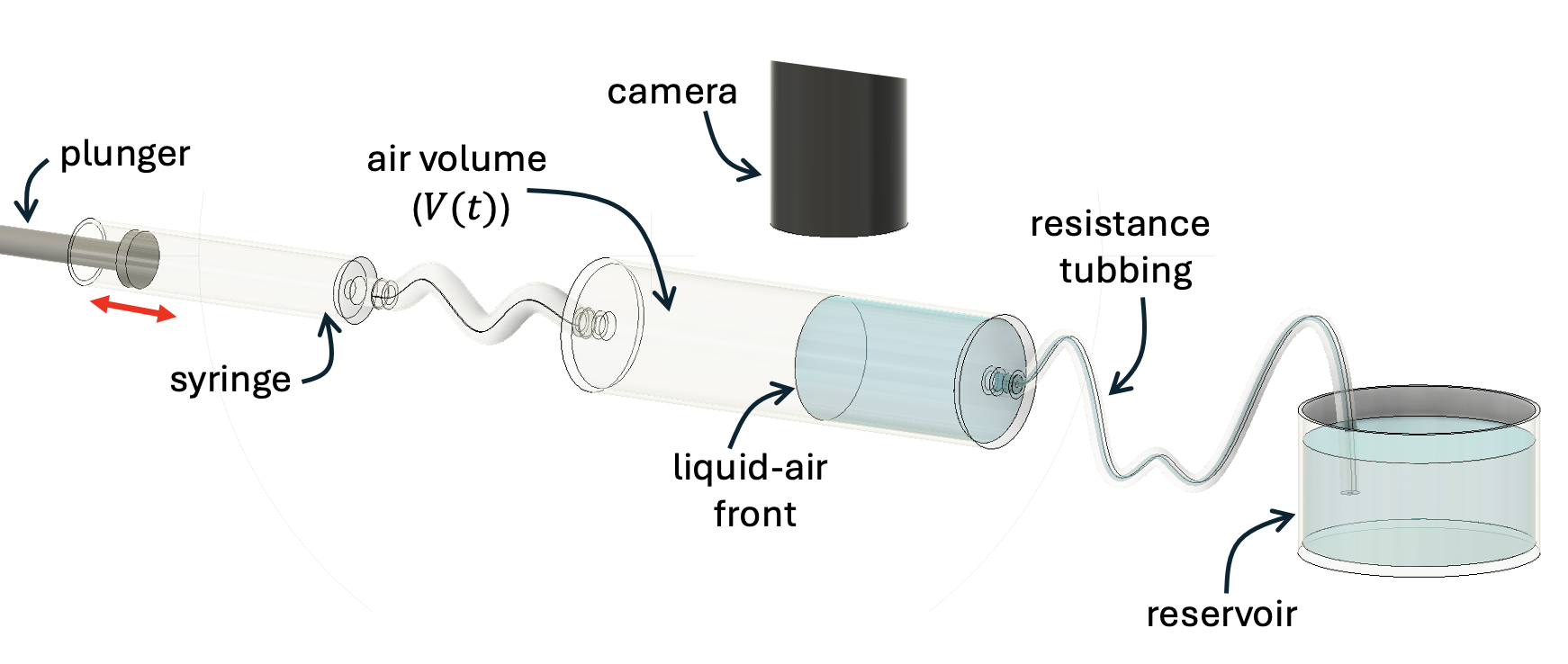}
    \caption{Schematic of the experimental setup for validating the damping model. The system consists of the oscillatory pump connected to a flexible tube, a glass observation tube containing the liquid-air interface, and a narrow capillary tube that provides dominant fluidic resistance, ending in a reservoir maintained at atmospheric pressure.}
    \label{fig:setup}
\end{figure}

The fluidic system, shown in Figure~\ref{fig:setup}, consists of the oscillatory pump connected to a flexible tube of length $L$ containing air, which determines the initial air volume $V_A$. This connects to a glass observation tube where the water-air front position is measured. The liquid section terminates in a narrow capillary tube (radius $r = 127$ \textmu m, variable length $l$) that provides the dominant fluidic resistance. The system ends in a water reservoir positioned to eliminate hydrostatic pressure effects by maintaining the same height as the liquid-air interface.

\end{multicols}
\subsection{Experimental setup}
\begin{multicols}{2}

Experimental parameters were systematically varied: frequency $\omega$ (0.37--8.38 s$^{-1}$) via motor voltage, fluidic resistance via capillary length $l$ (15--40 cm), and initial air volume $V_A$ ((4--6) $\cdot$ 10$^{-7}$ m$^3$) via flexible tube length $L$.

The system is placed under a high-magnification digital microscope camera (Retoo Digital Microscope, 2560 $\times$ 1920 pixels) that records both the syringe plunger and liquid-air front motion. When activated, the pump generates periodic sinusoidal compression and expansion of the air volume, which transmits motion to the liquid front through the compressible air phase.

\subsection{Data acquisition and calibration}
The camera records video at 30 fps, capturing the oscillatory movement of both the plunger and liquid front with high spatial resolution. ImageJ software \cite{imageJ} extracts position data in pixels on a frame-by-frame basis for both components. 

Spatial calibration was achieved using the known diameter of the glass observation tube. By measuring the tube diameter in pixels and knowing its actual physical dimensions, a precise pixel-to-length conversion factor was established. Volume displacements were then calculated using the cross-sectional geometry of the tube and the measured position changes.

The initial air volume $V_A$ was determined from the geometry of the system components: the syringe air volume, the flexible tube of length $L$, and the air-filled portion of the glass observation tube. The compliance $\mathcal{C}$ was calculated as $\mathcal{C} = V_A/K$, where $K$ is the bulk modulus of air, taken as the atmospheric pressure (101.3 kPa) under the assumptions of isothermal compression and small-strain.

The time constant $\tau$ for each experimental configuration was calculated using Equation (8), with values spanning from 8.7 to 17.5 seconds across the parameter space. These data were processed using MATLAB \cite{MATLAB} to convert pixel positions to physical displacements and generate comparative analysis between experimental results and theoretical predictions.

\section{Results and Discussion}

\begin{figure}
    \centering
    \includegraphics[width=\textwidth]{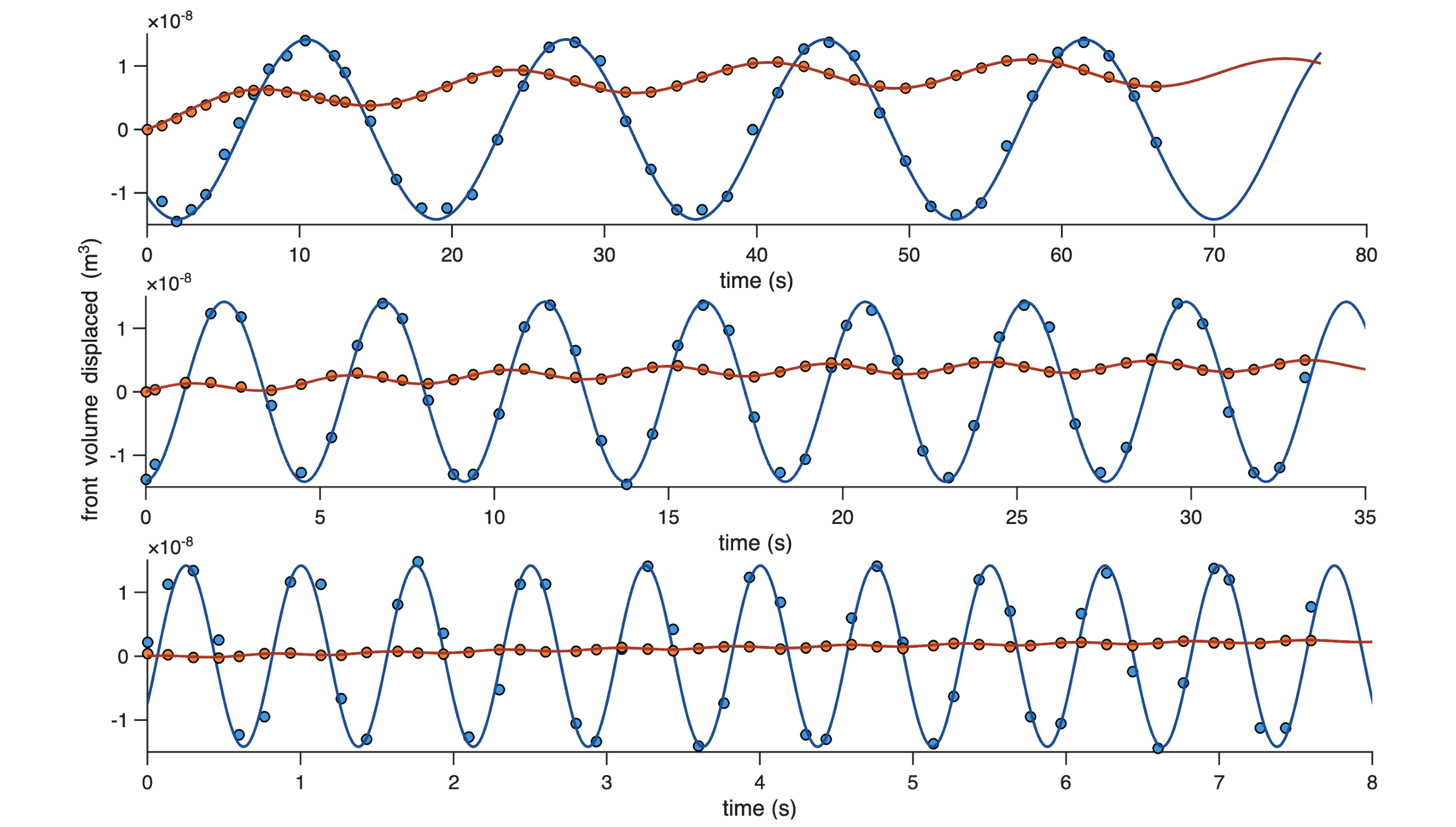}
    \caption{Experimental time-series of the volume displaced by the piston, $V_p(t)$ (blue, Eq.~\ref{eq:V_piston}), and the liquid front, $V(t)$ (orange, Eq.~\ref{eq:V_solution_final}), for three systems with varying $\tau\omega$. Experimental data for the liquid front are shown as points. System parameters: top: $\omega=0.37\ \mathrm{s}^{-1}$, $\mathcal{R}=2.94\times 10^{12}\ \mathrm{Pa\ s/m}^3$; middle: $\omega=1.37\ \mathrm{s}^{-1}$, $\mathcal{R}=1.96\times 10^{12}\ \mathrm{Pa\ s/m}^3$; bottom: $\omega=8.37\ \mathrm{s}^{-1}$, $\mathcal{R}=1.47\times 10^{12}\ \mathrm{Pa\ s/m}^3$; all with $\mathcal{C}=5.94\times 10^{-12}\ \mathrm{m}^3/\mathrm{Pa}$.}
    \label{fig:experiments}
\end{figure}

Figure \ref{fig:experiments} presents experimental data from three representative systems spanning different $\tau\omega$ regimes, providing a comprehensive validation of the theoretical model. The comparison between piston motion (blue, Eq.~\eqref{eq:V_piston}) and liquid-front response (orange, Eq.~\eqref{eq:V_solution_final}) reveals the full damping behavior predicted by the model. While the oscillation frequency remains identical between piston and front, confirming the system's linear character, three key modifications emerge: significant amplitude attenuation, substantial phase shift, and characteristic transient drift. The agreement between theoretical curves and experimental data across all $\tau\omega$ regimes demonstrates the predictive power of the model.

\end{multicols}
\begin{multicols}{2}

The progression from the top ($\tau\omega \approx 6.4$) to the bottom ($\tau\omega \approx 73$) panels illustrates the governing role of $\tau\omega$. At the lowest $\tau\omega$, the system responds quasi-statically with moderate damping. As $\tau\omega$ increases, the air compliance increasingly filters the oscillatory component, with the highest $\tau\omega$ case showing strong attenuation of the oscillatory signal. The consistent presence of transient drift across all regimes validates the model's ability to capture both steady-state and transient dynamics.

The selected $\tau\omega$ values (6.4 to 146) in all different seven experiments performed, were chosen to demonstrate the model's predictive capability across substantially different operating conditions, rather than to define system limits. Across this range, amplitude reductions vary from 85\% to 99.6\%, while phase shifts span 1.43 to 1.56 radians. This progression illustrates how increasing $\tau\omega$ enhances damping effects, with the highest $\tau\omega$ case showing near-complete attenuation of the oscillatory component. The consistent presence of transient drift across all systems further validates the model's completeness in capturing both steady-state and transient behaviors.

\begin{figure} [H]
    \centering
    \includegraphics[width=0.45\textwidth]{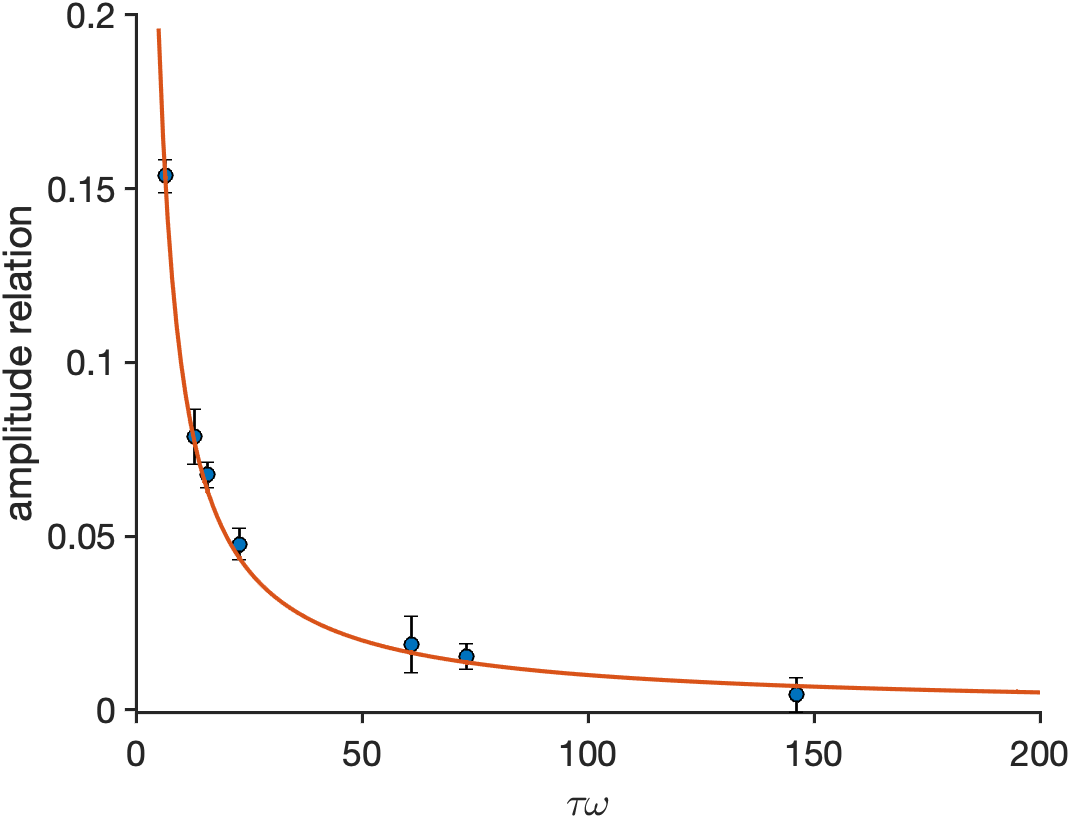}
    \caption{Amplitude relation between the plunger motion and the fluid front as a function of the $\tau \omega$ product for each experiment. The theoretical amplitude ratio obtained in equation \eqref{eq:amplitude_ratio} is plotted as a solid orange line.}
    \label{fig.amplitude}
\end{figure}

The physical mechanism underlying amplitude reduction stems from the fundamental competition between air compressibility and fluidic resistance. When fluid resistance dominates, the system responds by compressing the air volume rather than displacing liquid, effectively attenuating the transmitted motion. This energy distribution between compression work and flow work dictates the amplitude transmission ratio. The quantitative agreement shown in Figure \ref{fig.amplitude} demonstrates that this complex energy partitioning is perfectly captured by the dimensionless product $\tau\omega$, with experimental data following the theoretical prediction of Eq. \eqref{eq:amplitude_ratio} across the tested range.

\begin{figure}[H]
    \centering
    \includegraphics[width=0.45\textwidth]{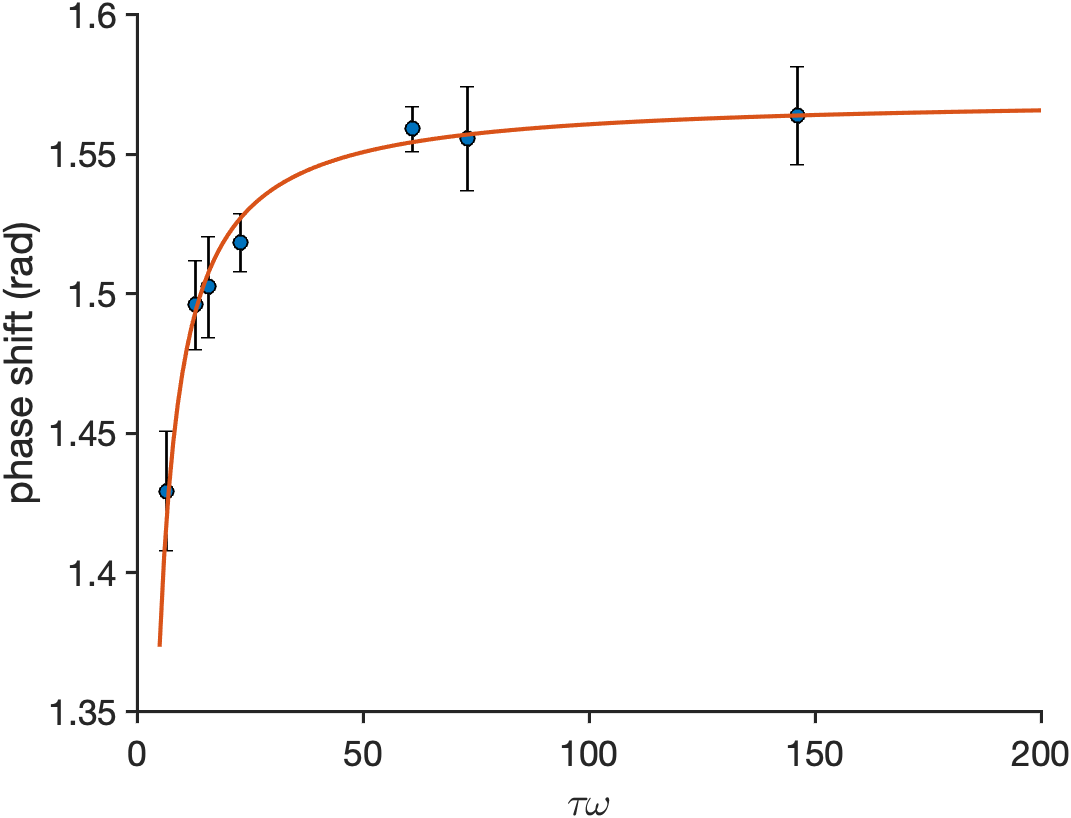}
    \caption{Phase shift between the movement of the plunger and the fluid front at the end of the $\tau \omega$ product for each experiment. The theoretical phase shift obtained in equation \eqref{eq:phase_shift} is plotted as a solid orange line.}
    \label{fig.phi}
\end{figure}

Phase shift measurements (Figure \ref{fig.phi}) provide additional validation of the theoretical framework. The observed phase lags follow the predicted arctangent dependence on $\tau\omega$, with minor stochastic variations attributable to measurement uncertainty in determining phase angles. Physically, these phase shifts manifest from the delayed transmission of motion through the compressible air volume, where energy storage in the compressed air introduces a temporal lag between piston motion and front response. The consistency across different $\tau\omega$ regimes confirms that the phase behavior is fundamentally governed by the same dimensionless parameter that controls amplitude reduction.

The transient drift observed in all experiments represents a particularly insightful validation of the model's completeness. This drift comprises two distinct components: one dependent on the initial piston position $\xi$, and another that persists even when $\xi=0$. The latter component, mathematically expressed as $\frac{V_0 \tau\omega e^{-t/\tau}}{\tau^2 \omega^2 + 1}$, just when $\xi=0$, represents a fundamental dynamic response where the system establishes its operating point through initial energy redistribution. This behavior, previously unreported in literature, emerges naturally from the first-principles derivation and is consistently observed across all experimental conditions.

The comprehensive validation across systematically varied air volumes ($V_A$), frequencies ($\omega$), and tube lengths ($l$) confirms the model's robustness and generality. By achieving $\tau\omega$ values from 6.4 to 146 through independent parameter variation, the experiments demonstrate that $\tau\omega$ serves as a universal governing parameter that encapsulates the essential physics of oscillatory damping in air-liquid systems.

\section{Conclusions}
\label{sec:conclusions}

This study establishes a comprehensive theoretical and experimental framework for understanding and harnessing the damping effects of a trapped air volume in oscillatory microfluidic flows. By deriving a mathematical model that fundamentally unites the elastic principles of a compressible air phase with the resistive principles of viscous liquid flow, I have moved beyond the empirical correlations that have limited previous studies. The model successfully captures the full dynamics of the liquid-air front, providing accurate, predictions of the amplitude reduction, phase shift, and previously unexplained transient mean motion.

The key insight from this work is the identification of the dimensionless product $\tau\omega$ as the fundamental parameter governing system behavior. This parameter, which represents the interplay between the system's inherent time constant (dictated by air compliance and fluidic resistance) and the driving frequency, provides a direct pathway for precise tuning of oscillatory flow characteristics. This experimental validation, using a custom-built syringe-pump system and image analysis, confirms the model's quantitative accuracy. The agreement is particularly strong for amplitude reduction, with all experimental data for the phase shift respecting the theoretical prediction within error margins.

The practical implications of this research are substantial. I transform the traditionally problematic presence of air from a source of unpredictability into a reliable and powerful design element. This model can be used as a predictive tool to design oscillatory microfluidic systems with tailored damping. By calculating the required $\tau\omega$ value for a desired amplitude ratio, designers can achieve precise flow stabilization and noise reduction by controlling the initial air volume, tubing geometry, or liquid viscosity, without resorting to complex active control mechanisms. This enables the design of systems that are both more precise and more robust.

Looking forward, this work provides the foundational framework for a new paradigm in analyzing microfluidic systems involving compressible phases. The methodology developed here can be directly applied to any non-stationary scenario with air and microfluidics, from pressure sensing to bubble-based logic. Future studies can now build upon this foundation to explore dynamics with non-Newtonian fluids, more complex channel geometries, or other types of compliance. In conclusion, this study provides both a fundamental understanding and a practical toolkit for leveraging air damping, paving the way for more predictable and high-performance oscillatory microfluidic systems.




\bibliography{rsc} 
\bibliographystyle{ieeetr} 
\end{multicols}
\end{document}